\documentclass[prd,amsfonts,aps,nofootinbib,notitlepage,12pt,superscriptaddress]{revtex4-1}

\pdfoutput=1
\usepackage{amsmath,amssymb}
\usepackage{graphicx}
\usepackage[utf8]{inputenc}
\usepackage{cancel}
\usepackage{slashed}
\usepackage{color}
\usepackage{array,multirow} 
\usepackage{subcaption}
\usepackage[normalem]{ulem}
\usepackage[colorlinks=true]{hyperref}
\usepackage{verbatim}

\newcommand{\Trh}{T_\text{rh}}

\newcommand{\ITM}{Institución Universitaria ITM, Facultad de Ciencias Exactas y Aplicadas, \\Calle 73 N° 76-354 vía el volador, Medellín, Colombia.}

\begin{document}

\title{ FIMP dark matter in the scotogenic model at low-reheating temperatures}

\author{Robinson Longas}
\email{robinsonlongas@itm.edu.co}
\affiliation{\ITM}
\author{Andrés Rivera}
\email{afelipe.rivera@udea.edu.co}
\affiliation{Instituto de Física, Universidad de Antioquia UdeA,
Calle 70 No. 52-21, Medellín, Colombia.}
\author{David Suarez}
\email{david.suarezr@udea.edu.co}
\affiliation{International Institute of Physics, Universidade Federal do Rio Grande do Norte, Campus Universit\'ario, Lagoa Nova, Natal-RN 59078-970, Brazil}
\affiliation{Departamento de F\'isica, Universidade Federal do Rio Grande do Norte, 59078-970, Natal, RN, Brasil}
\affiliation{Millennium Institute for Subatomic Physics at High-Energy Frontier (SAPHIR), Fernandez Concha 700, Santiago, Chile.}
\affiliation{Departamento de F\'isica, Facultad de Ciencias, Universidad de La Serena, Avenida Cisternas 1200, La Serena, Chile.}
\date{\today}

\begin{abstract}
The scotogenic model offers an attractive framework that explains light neutrino masses while accommodating a viable dark matter candidate. In this work, we study the phenomenology of singlet fermion dark matter, $N_1$, produced via the freeze-in mechanism characterized by an instantaneous low reheating temperature, $T_{\text{rh}}$. 
We explore the model's viable parameter space, incorporating current bounds from lepton flavor violation ($\mu \to e \gamma$) and direct detection limits. We show that low-reheating scenarios significantly reshape the viable region for $N_1$ dark matter, improving prospects for testing the scotogenic model in the FIMP scenario while keeping it consistent with current lepton flavor violation and cosmological bounds.
\end{abstract}

\maketitle

\section{Introduction}

Despite its remarkable experimental success, the Standard Model (SM) fails to account for several well-established observations. In particular, it does not explain the baryon asymmetry of the Universe (BAU)~\cite{AndreiDSakharov_1991}, lacks a viable dark matter (DM) candidate~\cite{1980ApJ...238..471R,Planck:2018vyg}, and provides no mechanism for generating neutrino masses~\cite{deSalas:2020pgw,Capozzi:2025wyn,Esteban:2024eli,ParticleDataGroup:2024cfk}. These shortcomings strongly motivate physics beyond the SM.

Among the proposed DM candidates, weakly interacting massive particles (WIMPs) remain the most extensively studied framework~\cite{Arcadi:2024ukq}. In this paradigm, the observed relic abundance arises from thermal freeze-out. Consequently, WIMPs can be probed through direct detection (DD)~\cite{Goodman:1984dc}, indirect detection (ID)~\cite{Fermi-LAT:2016uux,Lefranc:2016srp,IceCube:2016dgk,Albert:2016emp}, and collider searches~\cite{CMS:2016gsl}. However, the absence of any conclusive signal has placed increasingly stringent constraints on large regions of WIMP parameter space for several simplified models~\cite{Arcadi:2017kky}, motivating the exploration of alternative production mechanisms.

A compelling alternative is the freeze-in mechanism~\cite{Hall:2009bx}, in which DM is produced through extremely feeble interactions with the visible sector and never attains thermal equilibrium. The corresponding DM candidates, known as feebly interacting massive particles (FIMPs), are often considered challenging to test experimentally because of their suppressed couplings. Unlike freeze-out, which is largely governed by the thermal conditions near the decoupling temperature, freeze-in production can receive dominant contributions from much earlier epochs. Consequently, the final DM abundance can remain strongly sensitive to the maximum temperature reached during reheating, $T_{\text{rh}}$. In scenarios where the reheating temperature is low relative to the DM mass scale, the couplings required for freeze-in can be several orders of magnitude larger than in standard high-reheating scenarios, opening up prospects for testing the model via direct detection experiments~\cite{Hambye:2018dpi}.

On the other hand, an attractive framework that simultaneously addresses the DM and neutrino mass puzzles is the scotogenic model~\cite{Tao:1996vb,Ma:1998dn}, in which neutrino masses are generated radiatively and a stable DM candidate arises naturally from the symmetry structure of the model. In its minimal realization, the scotogenic model extends the SM by introducing an inert scalar doublet and three singlet fermions odd under a discrete $\mathbb{Z}_2$ symmetry. This symmetry simultaneously forbids tree-level neutrino masses, ensures the stability of the DM candidate, and allows one-loop neutrino masses~\cite{Restrepo:2021kpq,Longas:2023bvq,Agudelo:2024luc}.

In the scotogenic model, researchers have investigated FIMP DM assuming a standard high-reheating scenario~\cite{Molinaro:2014lfa}. More recently, researchers explored the impact of non-standard cosmological histories for WIMP DM~\cite{Roy:2025moo}, revealing substantial modifications to both relic-density predictions and the viable parameter space. Motivated by these findings, we investigate the consequences of instantaneous low-reheating scenarios on FIMP DM phenomenology within the scotogenic framework.

The paper is organized as follows: In \autoref{sec:model}, we provide an overview of the essential features of the scotogenic model. In \autoref{sec:Constraints}, we discuss the theoretical and phenomenological constraints imposed on the theory. In \autoref{sec:DM}, we examine the phenomenology of FIMP DM in the low-reheating regime. In \autoref{sec:num}, we present a numerical analysis of the parameter space. Finally, we draw our conclusions in \autoref{sec:Conclusions}.

\section{The model}
\label{sec:model}

The model under consideration is the scotogenic model, an extension of the SM that includes a DM candidate and generates Majorana neutrino masses~\cite{Tao:1996vb,Ma:1998dn}, radiatively generated at one loop through the exchange of neutral heavy scalars and Majorana fermions. The model has an exact discrete $\mathbb{Z}_2$ symmetry that guarantees DM stability. The SM field content is extended by three right-handed Majorana neutrinos, $N_i$ ($i=1,2,3$), which are singlets under the SM gauge group, and an inert scalar doublet, $\eta$ ~\cite{Longas:2015sxk,Branco:2011iw,LopezHonorez:2010eeh,LopezHonorez:2010tb}. $N_i$ and $\eta$ are odd under the $\mathbb{Z}_2$ symmetry, whereas all SM fields are even. The field content and quantum number assignments of the model are summarized in ~\autoref{tab:pickedsltn}.
\begin{table}[ht]
  \centering
  \begin{tabular}{|l|c|c|c|c|c|}\hline
    Field                       & Spin        & Generations   & $\mathrm{SU(2)_L}$     & $\mathrm{U(1)_Y}$ & $\mathbb{Z}_2$  \\ \hline
    $L_i$                       & $1/2$       & $3$           & $\mathbf{2}$  & $-1/2$                  & $+1$   \\  
    $e_i$                       & $1/2$       & $3$           & $\mathbf{1}$  & $1$                     & $+1$   \\  
    $N_j$                       & $1/2$       & $3$           & $\mathbf{1}$  & $0$                     & $-1$   \\ \hline 
    $\phi$                         & $0$           & $1$           & $\mathbf{2}$  & $+1/2$                  & $+1$   \\             
    $\eta$                         & $0$           & $1$           & $\mathbf{2}$  & $+1/2$                  & $-1$   \\ \hline             \end{tabular}
  \caption{Particle content and quantum numbers of the model.}
  \label{tab:pickedsltn}
\end{table}

The most general Lagrangian invariant under the symmetries of the model contains the following terms
\begin{equation}\label{eq:lagrangian}
\mathcal{L} =  \mathcal{L}_{\text{SM}} + \bar{N}_i i \partial \! \! \! / N_i - \frac{m_i}{2} \bar{N}_i^c N_i + \left(D_{\mu}\eta\right)^{\dagger}{D}^{\mu}\eta + \left(y_{\alpha i} \bar{L}_\alpha \tilde{\eta} N_i + \text{h.c.} \right)-V\left(\phi,\eta\right)\,,
\end{equation}
where $y_{\alpha i}$ are the Yukawa couplings and $\tilde{\eta}=i\sigma_2\eta$. $V(\phi,\eta)$ is the scalar potential 
\begin{align}
V(\phi,\eta) = &\,m_1^2 \phi^\dagger \phi + m_2^2 \eta^\dagger \eta + \frac{\lambda_1}{2} (\phi^\dagger \phi)^2 + \frac{\lambda_2}{2} (\eta^\dagger \eta)^2\nonumber \\
&+ \lambda_3 (\phi^\dagger \phi)(\eta^\dagger \eta) + \lambda_4 (\phi^\dagger \eta)(\eta^\dagger \phi) + \frac{\lambda_5}{2} \left[ \left(\phi^\dagger \eta\right)^2 + \mathrm{h.c.}\right]\,.    
\end{align}
Because the inert doublet $\eta$ is odd under the $\mathbb{Z}_2$ symmetry, it does not acquire a vacuum expectation value and lacks Yukawa couplings to SM quarks; tree-level flavor changing neutral currents (FCNCs) mediated by neutral scalars are absent. The doublets $\phi$, $\eta$ can be written as 
\begin{align}
\label{eq:doubletes}
\eta =  \begin{pmatrix}
 H^{+}  \\
 \frac{H^0 + i A^0}{\sqrt{2}}
\end{pmatrix}\;,\;
\phi =  \begin{pmatrix}
 G^+  \\
 \frac{h + v + i G^0}{\sqrt{2}}
\end{pmatrix}\;,
\end{align}
and after electroweak symmetry breaking, $\left \langle \phi \right \rangle = \left( 0, v/\sqrt{2} \right)^T$ with $v\approx 246\, \mathrm{GeV}$, the scalar mass eigenstates of the model consist of the scalar Higgs $h$ with mass $m_{h}=125\, \text{GeV}$ and the $ \mathbb {Z} _ 2$-odd states $H^0$, $A^0$ and $H^{\pm}$ with masses
\begin{align}
\label{eq:bosomspectrum}
m_{H^0}^2 =~ m_2^2 + v^2\frac{\lambda_3 + \lambda_4 + \lambda_5}{2},\;\;
m_{A^0}^2 =~ m_2^2 + v^2\frac{\lambda_3 + \lambda_4 - \lambda_5}{2},\;\;
m_{H^\pm}^2 =  m_2^2 + v^2\frac{\lambda_3}{2}\,.
\end{align}
As in~\cite{Tao:1996vb}, the mass splitting between the CP-odd and CP-even bosons is controlled by the scalar parameter $\lambda_5$, which enters in neutrino mass generation. We define $\lambda_L \equiv (\lambda_3 + \lambda_4 + \lambda_5)/2$ and choose the free parameters in the potential as $\{ m_{H^0}^2, m_{A^0}^2, m_{H^{\pm}}^2, \lambda_L, \lambda_2 \}$.  The remaining parameters are
\begin{align}
\label{eq:non_freePotentialParameters}
m_2^2 & = m_{H^0}^2 - v^2\lambda_L \,,\; \lambda_3 = \frac{2}{v^2}\left( m_{ H^{\pm}}^2 -m_{H^0}^2\right) +2\lambda_L\,,\nonumber\\
 \lambda_4 &= \frac{1}{v^2}\left(  m_{H^0}^2 + m_{A^0}^2 - 2m_{ H^{\pm}}^2  \right)\,,\;
 \lambda_5 = \frac{1}{v^2}\left( m_{H^0}^2 - m_{ A^0}^2 \right)\,.
\end{align}
In this model, the mass term for heavy neutrinos violates total lepton number by 2 units, allowing Majorana neutrino masses to be generated via one-loop diagrams mediated by the odd neutral scalars. The resulting mass matrix is given by 
\begin{align}\label{eq:neutrinomasses}
M^{\nu}_{\alpha\beta} = \sum_{k=1}^3\frac{y_{\alpha{k}}y_{\beta{k}}}{16\pi^2}m_k \left[ \frac{m_{H^0}^2}{m_{H^0}^2 - m_k^2} \ln\left( \frac{m^2_{H^0}}{M_k^2}\right) - \frac{m_{A^0}^2}{m_{A^0}^2 - m_k^2} \ln\left( \frac{m_{A^0}^2}{m_k^2}\right) \right]\,,
\end{align}
which in matrix form is $ M^{\nu} = y \Lambda y^T$, with $\Lambda=\operatorname{diag}(\Lambda_1, \Lambda_2, \Lambda_3)$ and
\begin{align}\label{eq:neutrinomasses3}
\Lambda_k  = \frac{m_k}{16\pi^2} \left[ \frac{m_{H^0}^2}{m_{H^0}^2 - m_k^2} \ln\left( \frac{m_{H^0}^2}{m_k^2}\right) - \frac{m_{A^0}^2}{m_{A^0}^2 - m_k^2} \ln\left( \frac{m_{A^0}^2}{m_k^2}\right) \right]\,.
\end{align}

Notice that when $m_{H^0} = m_{A^0}$ ($\lambda_5 = 0 $), the Majorana neutrino mass matrix vanishes. In the limit of small $\lambda_5$, the neutrino mass matrix is proportional to $\lambda_5$, naturally explaining the smallness of neutrino masses. 

\section{Constraints}
\label{sec:Constraints}

To ensure the model's theoretical consistency and compatibility with current experimental data, we impose the following constraints throughout our numerical analysis.

\subsection{Theoretical Constraints}

\begin{itemize}
    
\item \textbf{Vacuum stability:} We demand that the model satisfy the following criteria for the quartic couplings of the scalar potential to guarantee the stability of the vacuum~\cite{Belyaev:2016lok}:
\begin{align}
\lambda_1>0,\;\;  \lambda_2>0,\;\;\lambda_3+\sqrt{\lambda_1\lambda_2}>0,\;\;
\sqrt{\lambda_1 \lambda_2}+\lambda_3+\lambda_4-|\lambda_5|\geq 0 \,.
\end{align}

\item \textbf{Perturbativity:} To ensure the validity of perturbation theory, we require all scalar couplings to remain in the perturbative regime~\cite{Belanger:2022qxt,Garcia-Cely:2015khw}:
\begin{align}
\lambda_2<\frac{2\pi}{3},\;\;|\lambda_3|<4\pi,\;\;|\lambda_4|<4\pi,\;\;|\lambda_5|<4\pi,\;\;|\lambda_3+\lambda_4\pm\lambda_5|<4\pi\,.
\end{align}

\item \textbf{Perturbativity Unitarity:} Tree-level perturbative unitarity requires all eigenvalues of the scalar $2\to2$ scattering matrix to satisfy the unitarity bounds~\cite{Arhrib:2012ia}. We impose:
\begin{align}
&\left|\lambda_3+3\lambda_4\pm{2}\lambda_5\right|<4\pi,\;\;
\left|-\lambda_1-\lambda_2\pm\sqrt{(\lambda_1-\lambda_2)^2+4\lambda^2_4}\right|<16\pi\,,\\
&\left|-3\lambda_1-3\lambda_2\pm\sqrt{9(\lambda_1-\lambda_2)^2+4(2\lambda_3+\lambda_4)^2}\right|<16\pi\,,\\
&\left|-\lambda_1-\lambda_2\pm\sqrt{(\lambda_1-\lambda_2)^2+4\lambda^2_5}\right|<16\pi\,.
\end{align}

\item \textbf{Inert vacuum condition:} We require the inert vacuum to be the global minimum of the scalar potential~\cite{Ginzburg:2010wa}, which guarantees that the $\mathbb{Z}_2$ symmetry remains unbroken and the DM candidate is stable
\begin{equation}
m^2_2 > \sqrt{\frac{\lambda_1}{\lambda_2}} m^2_1 \times 
\begin{cases} 
f_m & \text{if } |f_m| < 1 \,, \\ 
1 & \text{if } |f_m| \ge 1 \,,
\end{cases}
\end{equation}
with $f_m=\dfrac{\lambda_L}{\sqrt{\lambda_1\lambda_2}} $. 
We additionally require $\lambda_4-|\lambda_5|<4\pi$ to avoid charge-breaking vacua ~\cite{Belyaev:2016lok}.
\end{itemize} 

\subsection{Phenomenological constraints}

\begin{itemize}

\item \textbf{Charged Lepton Flavor Violation:} The Yukawa couplings responsible for neutrino mass generation induce charged lepton flavor violating (CLFV) processes at one loop. We require $\mu\to e\gamma$ to satisfy the current experimental bound $\mathcal{B}(\mu\to e\gamma) <1.5\times10^{-13}$~\cite{MEGII:2025gzr}. This observable provides one of the strongest constraints on the model's Yukawa sector. In the scotogenic model, the branching ratio for $\mu\to{e}\gamma$ is~\cite{Toma:2013zsa,Vicente:2014wga}
\begin{equation}
\label{eq:muegamma}
\mathcal{B}(\mu\to{e}\gamma)=\frac{3(4\pi)^3\alpha_{\text{EM}}}{4G^2_F}|A_D|^2\mathcal{B}(\mu\to{e}\nu_{\mu}\overline{\nu}_e)\,,
\end{equation}
where $\alpha_{\text{EM}}=e^2/4\pi$ and the explicit form of $A_D$ is in Ref.~\cite{Toma:2013zsa}.

\item \textbf{Lifetime of Unstable Species:} Long-lived particles may alter the successful predictions of Big Bang Nucleosynthesis (BBN). We require all unstable states to decay before the onset of BBN to avoid this possibility~\cite{Allahverdi:2020bys}
\begin{equation}
\tau=\frac{1}{\Gamma}<1\,\text{s}\,.
\end{equation}

In particular, the decay rate $N_j\to{N}_i\overline{l}_{\alpha}l_{\beta}$ is~\cite{Molinaro:2014lfa}
\begin{align}
\Gamma(N_j\to{N}_i\overline{l}_{\alpha}l_{\beta})=\frac{m^5_j}{6144\,\pi^3m^4_{H^{\pm}}}\left(|y_{\beta{i}}|^2|y_{\alpha{3}}|^2+|y_{\alpha{i}}|^2|y_{\beta{3}}|^2\right) \,.   
\end{align}

\item \textbf{Collider Bounds on Scalar  Masses:} Direct searches at LEP impose a lower bound on the charged scalar mass. In particular, searches for $e^+e^- \rightarrow H^+H^-$ exclude charged scalar masses below $m_{H^\pm}<79.3~{\rm GeV}$~\cite{ALEPH:2002ftu}, provided that the decay channel $H^\pm\to A^0W^\pm$ is kinematically forbidden. In addition, measurements of the $Z$-boson width require $m_{H^0}+m_{A^0}>m_Z$~\cite{Lundstrom:2008ai}. Furthermore, if $|m_{H^0}-m_{A^0}|>8$ GeV, the CP-odd and CP-even scalars are excluded for masses $m_{A^0} < 80 $ GeV and $m_{H^0}<100$ GeV~\cite{Lundstrom:2008ai}.

\item \textbf{Invisible Higgs Decay:} The invisible Higgs decays into two lighter scalars are kinematically forbidden if we consider $m_{A^0} > m_h/2$ and $m_{H^0} > m_h/2$. 

\item  \textbf{Oblique corrections:} The new doublet contributes at one-loop level to the self-energies of the SM gauge bosons. DM candidates are electroweak singlets and do not contribute. We use the Peskin-Takeuchi oblique parameters $S$, $T$, and $U$ to quantify contributions
\begin{subequations}
\begin{eqnarray}
S|_{U=0} &=& 0.04\pm0.08 \; ,
\\
T|_{U=0} &=& 0.08\pm0.07 \; ,
\end{eqnarray}
\end{subequations}
with a correlation coefficient of 0.92~\cite{Haller:2018nnx}. 

\end{itemize}

\section{Dark matter}
\label{sec:DM}

The DM candidate is the lightest $\mathbb{Z}_2$  odd state, which can be either the CP-even scalar $H^0$ ($\lambda_5<0$), the CP-odd scalar $A^0$ ($\lambda_5>0$), or the lightest fermion state $N_1$. The WIMP phenomenology of the models has been studied extensively; see, for instance, Refs. \cite{Kubo:2006yx,AristizabalSierra:2008cnr,Roy:2025moo,Kashiwase:2012xd,Klasen:2013jpa,Chakrabarty:2015yia,LopezHonorez:2006gr}. In this work, we consider $N_1\equiv N$ as the DM candidate, namely, it is the lightest particle odd under the $\mathbb{Z}_2$  symmetry, in other words, $m_1\equiv m_N < \{m_2, m_3, m_{H^0}, m_{A^0}, m_{H^{\pm}}\}$. We study its relic abundance through non-thermal mechanisms: freeze-in \cite{Hall:2009bx} and the thermal superWIMP mechanism~\cite{Feng:2003xh}, in which heavier $\mathbb{Z}_2$-odd particles freeze out and subsequently decay into $N$.

\subsection{Freeze-in production}
\label{sec:FIMP}

In the FIMP scenario \cite{Hall:2009bx}, the DM candidate interacts so feebly with the SM sector that it never reaches thermal equilibrium with the SM plasma. Consequently, its relic abundance arises from non-thermal production.  

The DM candidate $N$ interacts with SM leptons through the Yukawa couplings $y_{\alpha 1}$. 
Its production is dominated by the two-body decays of the heavy neutral and charged scalars, 
whereas $2\to 2$ scatterings in the $t$-channel are subdominant as they scale as $\mathcal{O}(y_{\alpha 1}^4)$  compared to the $\mathcal{O}(y_{\alpha 1}^2)$ scaling of scalar decays~\cite{Molinaro:2014lfa}. The Boltzmann equation governs the time evolution of the DM number density: 
\begin{align}\label{eq:boltzmaneq1}
    s(T) T H(T) \frac{d Y_{N}}{d T} =  -\sum_{\chi= H^0,A^0,H^\pm}\, \frac{g_{\chi} m^2_{\chi} T}{2\pi^2}\, K_1(m_{\chi}/T)\, \Gamma(\chi \to L N )\,,
\end{align}
where $Y_N \equiv n_N/s$ is the DM yield, $s(T) = (2\pi^2 / 45) g_s T^3$ is the entropy density, and $H(T) = 1.66 \sqrt{g_*} T^2 / M_P$ is the Hubble parameter, with $M_P \approx 1.22 \times 10^{19}~\text{GeV}$ being the Planck mass.  The internal degrees of freedom are $g_{\chi}=2$ for $\chi=H^{\pm}$ and $g_{\chi}=1$ for $\chi=\{H^0, A^0\}$. The decay  widths are given by~\cite{Molinaro:2014lfa}
\begin{align} \label{eq:decayrate}
    \Gamma(H^0/A^0 \to \nu_{\alpha} N) &=   \frac{|y_{\alpha 1}|^2}{32\pi} m_{H^0/A^0} \left( 1 - \frac{m_N^2}{m_{H^0/A^0}^2} \right)^2\,, \\
    \Gamma(H^{\pm} \to l_{\alpha} N) &=   \frac{|y_{\alpha 1}|^2}{16\pi} m_{H^{\pm}} \left( 1 - \frac{m_N^2}{m_{H^{\pm}}^2} \right)^2\,. \label{eq:decayratee}
\end{align}
The corresponding three body decay, $N_{2,3} \rightarrow N\overline{\ell}\ell$ gives a negligible contribution compared with Eq.~\eqref{eq:decayrate} and Eq.~\eqref{eq:decayratee} \cite{Molinaro:2014lfa}.

Assuming instantaneous reheating and negligible DM production during the reheating epoch, we integrate Eq.~\eqref{eq:boltzmaneq1} from the reheating temperature $\Trh$ down to the current CMB temperature $T_0 \sim 2.7~\text{K}$:
\begin{align}\label{eq:boltzman_instantanea1}
    Y_N(T_0) \sim \sum_{\chi}\, \frac{45\,g_{\chi}}{4\pi^4 \, (1.66) \,g_s \sqrt{g_*}} \, m^2_{\chi} M_P \, \Gamma(\chi \to L N ) \int_{\Trh}^{T_0} dT' \, \frac{K_1(m_{\chi}/T')}{T'^5 }\,.
\end{align}
By introducing the dimensionless variable $x \equiv m_{\chi}/T$, the yield at present day simplifies to
\begin{align}\label{eq:boltzman_instantanea2}
    Y_N(T_0) \sim \sum_{\chi}\, \frac{6.77\,g_{\chi}}{\pi^4 \,g_s \sqrt{g_*}} \, \frac{M_P \Gamma(\chi \to L N)}{m^2_{\chi}} \int_{m_{\chi}/\Trh}^{m_{\chi}/T_0} dx \, x^3\, K_1(x) \,.
\end{align}
The final DM relic density is then related to the yield via
\begin{align}\label{eq:abundance1}
    \Omega_N h^2 \approx 2.7\times 10^8 \left(\frac{m_N}{\mathrm{GeV}}\right) Y_N(T_0)\,.
\end{align}

Depending on the hierarchy between the scalar masses and the reheating temperature, two distinct regimes emerge:

\subsubsection*{High-Temperature Reheating Regime (\texorpdfstring{$\Trh \gg m_{\chi}$}{Trh >> mchi})}
In the standard high-reheating scenario, the lower integration limit in Eq.~\eqref{eq:boltzman_instantanea2} vanishes ($m_\chi / \Trh \to 0$), leading to $\int_0^\infty dx \, x^3 K_1(x) = 3\pi/2$. Thus, the yield reduces to
\begin{align}\label{eq:boltzman_instantanea4}
    Y_N(T_0) \sim \sum_{\chi}\, \frac{6.77\,g_{\chi}}{\pi^4 \,g_s \sqrt{g_*}} \, \frac{M_P \, \Gamma(\chi \to L N )}{m^2_{\chi}} \left(\frac{3\pi}{2}\right) \,.
\end{align}
To gain analytical insight, we consider a compressed scalar spectrum $m_{A^0} = m_{H^0} = m_{H^\pm} \equiv n m_N$ (with $n > 2$), a single dominant Yukawa coupling $y_1 \equiv |y_{\alpha 1}|$, and effective relativistic degrees of freedom $g_s \approx 3.93$ and $g_* \approx 100$. Under these approximations, the yield evaluates to
\begin{align}\label{eq:Yaproximadoalto}
    Y_N(T_0) \sim 6.07 \times 10^{15} \, \frac{\left( n^2 - 1\right)^2}{n} \left( \frac{\mathrm{GeV}}{m_N}\right) y_1^2\,,
\end{align}
which yields a relic abundance independent of the dark matter mass:
\begin{align}\label{eq:abundance1altaT}
    \Omega_N h^2 \sim 1.64 \times 10^{24} \, \frac{\left(n^2-1\right)^2}{n} \, y_1^2\,.
\end{align}
Matching the observed Planck relic density $\Omega_N h^2 \approx 0.12$ fixes the required Yukawa coupling to
\begin{align}
    y_1 \sim 2.70 \times 10^{-13} \, \frac{\sqrt{n}}{n^2-1}\,.
\end{align}

\subsubsection*{Low-Temperature Reheating Regime (\texorpdfstring{$\Trh \ll m_{\chi}$}{Trh << mchi})}
If the reheating temperature is lower than the decaying scalar masses, production from the tail of the thermal distribution becomes Boltzmann suppressed. Using the asymptotic expansion of the Bessel integral for large lower limits, the yield simplifies to
\begin{align}\label{eq:boltzman_instantanea5}
    Y_N(T_0) \sim \sum_{\chi}\, \frac{6.77\,g_{\chi}}{\pi^4 \,g_s \sqrt{g_*}} \, \frac{M_P \, \Gamma(\chi \to L N ) }{m^2_{\chi}} \sqrt{\frac{\pi}{2}} \left(\frac{m_{\chi}}{\Trh} \right)^{5/2} e^{-m_{\chi}/\Trh}\,.
\end{align}
Using the same compressed spectrum assumptions, the yield and abundance become
\begin{align}\label{eq:Yaproximadobajo}
    Y_N(T_0) &\sim 1.61 \times 10^{15} \, \frac{\left( n^2 - 1\right)^2}{n} \left( \frac{\mathrm{GeV}}{m_N}\right) y_1^2 \left(\frac{n \, m_N}{\Trh}\right)^{5/2} e^{-\frac{n m_N}{\Trh}}\,, \\
\label{eq:abundancebajaT}
    \Omega_N h^2 &\sim 4.36 \times 10^{23} \, \frac{\left(n^2-1\right)^2}{n} \, y_1^2 \left(\frac{n \, m_N}{\Trh}\right)^{5/2} e^{-\frac{n m_N}{\Trh}}\,.
\end{align}
To reproduce $\Omega_N h^2 \approx 0.12$, the required Yukawa coupling must be enlarged to compensate for the exponential suppression:
\begin{align}\label{eq:y_lowT}
    y_1 \sim 5.24 \times 10^{-13} \, \frac{\sqrt{n}}{n^2-1} \left(\frac{n \, m_N}{\Trh}\right)^{-5/4} e^{\frac{n m_N}{2\Trh}}\,.
\end{align}
\begin{figure}[!htbp]
    \def\sepf{0.496}
    \centering
    \includegraphics[width=\sepf\columnwidth]{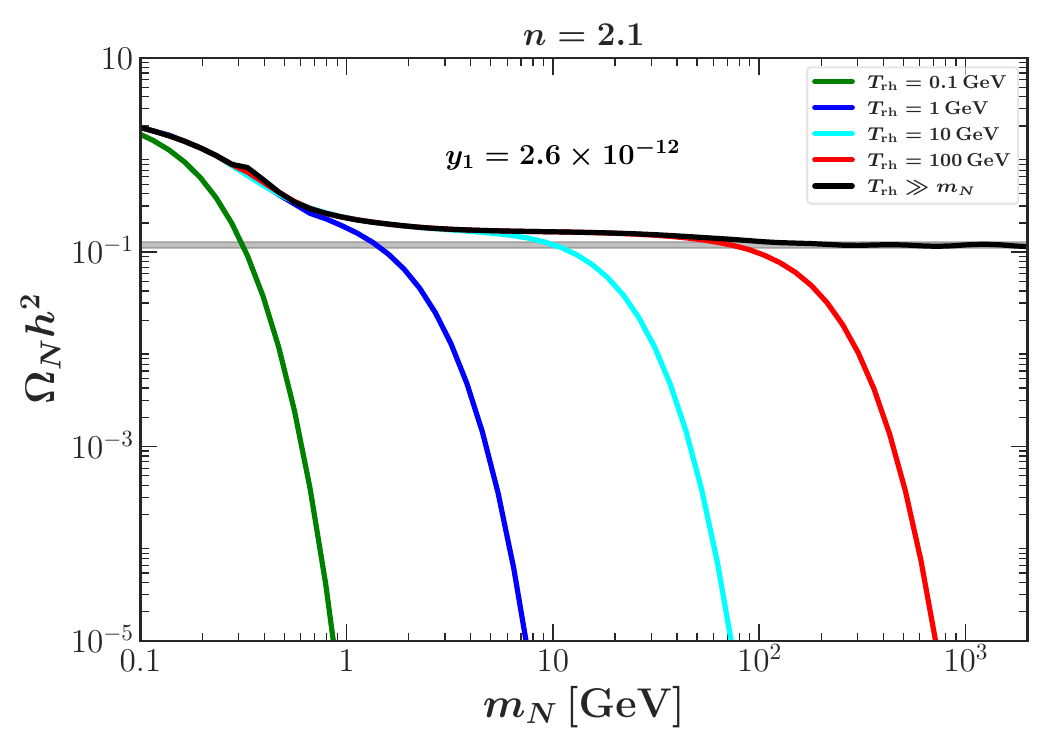}
    \includegraphics[width=\sepf\columnwidth]{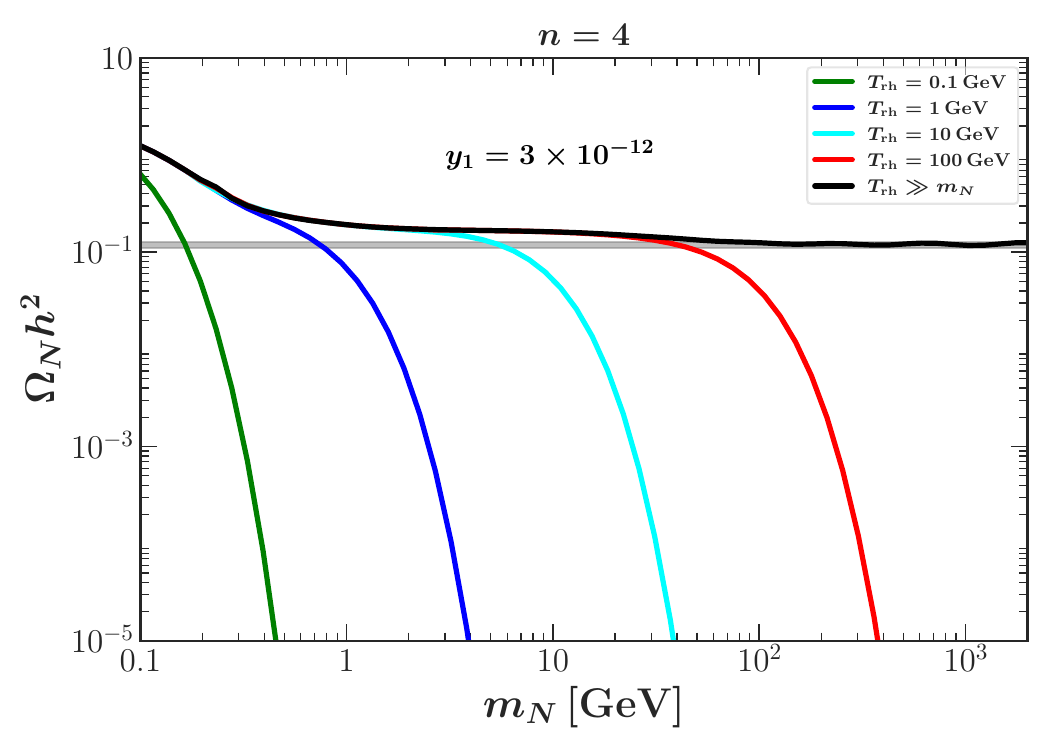}
    \caption{DM relic abundance as a function of the DM mass for different reheating temperatures. We have taken $m_{A^0} = nM_N = m_{H^0} = m_{H^\pm}$. } 
     \label{fig:relic_density}
\end{figure} 

\autoref{fig:relic_density} shows the DM relic density as a function of $m_N$ for various reheating temperatures. In the high reheating temperature regime (black curves), $\Omega_N h^2$ is flat with respect to $m_N$, reproducing the observed relic abundance for $y_1 \sim 10^{-12}$, consistent with our analytical estimates. Conversely, for $\Trh < m_\chi$ (colored curves), the severe Boltzmann suppression reduces the final relic density for heavier DM masses, demanding larger Yukawa couplings, $y_1$, to satisfy Planck bounds. 
However, $y_1$ cannot be increased arbitrarily because the freeze-in mechanism remains valid only as long as $N$ never enters thermal equilibrium with the SM bath, requiring
\begin{align}\label{eq:equilibrumconditions}
    \Gamma_{\text{eff}} \sim \sum_{\chi} \frac{n_{\chi}^{\rm eq}}{n_N^{\rm eq}}\,\Gamma(\chi \to L N) \lesssim H(T)\,.
\end{align}

\subsection{SuperWIMP}
\label{sec:superWIMP}

DM can also be produced through the super-WIMP mechanism, in which the next-to-lightest odd particle (NLOP) freezes out and subsequently decays into the DM particle. The NLOP remains in thermal equilibrium with the SM plasma and the other $\mathbb{Z}_2$-odd particles, while the field $N$ never reaches thermal equilibrium. 
However, as the Universe cools, the NLOP interaction rate falls below the Hubble expansion rate, leading to freeze-out. In this scenario, the DM relic abundance is given by~\cite{Molinaro:2014lfa, Guo:2025xmz}
\begin{equation}
    \Omega_N^{\text{superWIMP}} = \dfrac{m_N}{m_{\text{NLOP}}}\Omega_{\text{NLOP}}^{\text{freeze-out}}\,.
\end{equation}
Although in this model, there are two possibilities for the NLOP: the $N_2$ fermion or the lightest scalar $H^0(A^0)$, we only consider $N_2$ as the NLOP; otherwise, for scalars as NLOPs, the DM relic density is only obtained for masses above $600$ GeV~\cite{Molinaro:2014lfa}. From now on, we denote the NLOP as $N_2$.

$N_2$ decay into DM via $N_2\to\,N l\,\overline{l}$, which must occur after $N_2$ freeze-out, requiring $\Gamma(N_2\to\,N l\,\overline{l}) \lesssim H(T)$. This requirement leads to an upper bound~\cite{Molinaro:2014lfa}
\begin{equation}
\label{eq:n2ton1ll}
y_1\,y_2 \lesssim 2\times 10^{-6}\left(\dfrac{m_{\chi}}{1\text{TeV}}\right)\left(\dfrac{1\text{TeV}}{m_2}\right)^{3/2}\,,
\end{equation}
where $M_2$ is the mass of $N_2$ and we use the abbreviation $y_{\alpha i}=y_i$. 
On the other hand, to avoid disturbing the BBN, the lifetime of $N_2$ should be less than about $1\, \text{s}$. This condition yields a lower bound~\cite{Molinaro:2014lfa}
\begin{equation}
\label{eq:y1y2_BBN}
y_1\,y_2 \gtrsim 3\times 10^{-12}\left(\dfrac{m_\chi}{1\text{TeV}}\right)^2\left(\dfrac{1\text{TeV}}{m_2}\right)^{5/2}\,.
\end{equation}
Notice the interplay between Yukawa couplings $y_1$ and $y_2$ in the last two equations. They determine the success of the super-WIMP production mechanism. Moreover, $y_1$ governs the freeze-in production of $N$. 
In addition, the Yukawa couplings $y_i$ also appear in the Casas-Ibarra parametrization of the neutrino Yukawa matrix~\cite{Casas:2001sr}, which typically requires $y_2\,y_3\geq 10^{-6}$ to reproduce the observed neutrino scale~\cite{Molinaro:2014lfa}. Finally, notice that the same Yukawa couplings also contribute to CLFV processes, such as $\mu\to e\gamma$ computed in Eq.~\eqref{eq:muegamma}. Therefore, the interplay among neutrino masses, CLFV processes, and BBN is susceptible to the superWIMP mechanism.

\subsection{Testability}

In the standard freeze-in scenario with high reheating temperatures, the Yukawa couplings associated with DM production are extremely small, making the model hard to test experimentally. However, as discussed above, low reheating temperatures induce a Boltzmann suppression on the relic abundance. Larger Yukawa couplings are required to compensate for this suppression and match the observed DM density, bringing the model within reach of various experimental searches. In particular, the Yukawa interactions responsible for DM production mediate elastic scattering with both electrons and nucleons, giving rise to potential signals in direct detection experiments.

\begin{itemize}

\item \textbf{DM--Electron Scattering:} The interaction between the DM particle $N$ and electrons is mediated by the charged scalar $H^\pm$. In the limit $m_{H^\pm}\gg m_e,m_N$, the elastic scattering cross section renders
\begin{align}
\sigma_{e}\simeq\frac{y^4_1\mu_{eN}^{2}}{8\pi{m}_{H^\pm}^{4}},
\end{align}
where $y_1$ denotes the relevant Yukawa coupling and $\mu_{eN}=m_e m_N/(m_e+m_N)$ is the reduced mass of the electron--DM system. The cross section scales as $y^4_1$ and is strongly suppressed by the fourth power of the charged-scalar mass. Consequently, sizable signals require relatively light mediators or large Yukawa couplings. Since low-temperature reheating typically demands larger couplings to reproduce the observed relic abundance, future low-threshold experiments sensitive to electron recoils may probe part of the viable parameter space.

\item \textbf{Direct Detection via Nuclear Recoils:} The Yukawa interactions also induce effective DM-quark couplings through loop diagrams involving the inert scalar sector. These interactions generate spin-independent (SI) scattering off nuclei. The SI cross section is~\cite{Ibarra:2016dlb}
\begin{equation}
\sigma_{\rm SI}=\frac{4}{\pi}\mu_{pN}^2m_p^2\left(\frac{\Lambda_q}{m_q}\right)^2f_p^2\,,
\end{equation}

where $\mu_{PN}=m_P m_N/(m_P+m_N)$ is the reduced mass for the DM-proton system, $m_N$ denotes the DM mass, $m_p$ is the proton mass, $m_q$ is the quark mass, and $f_p$ is the proton form factor. The effective coupling $\Lambda_q$, generated at one loop level, depends on the masses of the inert scalars and the Yukawa couplings of the model. Ref.~\cite{Ibarra:2016dlb}.

\end{itemize}

Because SI interaction arises at loop level, the corresponding scattering rates are typically suppressed. Nevertheless, the enhanced Yukawa couplings required in low-temperature reheating scenarios can significantly increase these cross sections. Consequently, present and future direct-detection experiments may provide complementary tests of the parameter space that reproduces the observed DM relic abundance through freeze-in production.

\section{Numerical results}
\label{sec:num}

We consider the scalar couplings $(\lambda_2,\lambda_L)$ and the masses $(m_{H^0},m_{A^0},m_{H^\pm},m_N)$ to be free parameters of the model. Since $\lambda_2$ only controls the quartic self-interactions of the inert doublet and has a negligible impact on the DM phenomenology, we leave it unconstrained. Throughout the analysis, we fix $\lambda_L=1$, which maximizes the spin-independent DM-nucleon scattering cross section.

For the DM mass, we scan the range $ 1\,\mathrm{GeV} \leq m_N \leq 10^{4}\,\mathrm{GeV}$,
while the scalar masses are parameterized as
\begin{equation}
\label{eq:scan-parameters}
m_{A^0}=nM_N,\qquad
m_{H^0}=m_{A^0}+\delta,\qquad
m_{H^\pm}=m_{H^0} + 80\,\mathrm{GeV},
\end{equation}
where $n$ characterizes the dark-sector mass hierarchy and $\delta$ determines the splitting between the neutral CP-odd and CP-even scalars, which controls the $\lambda_5$ parameter. We consider $\delta = 10^{-5}\,\text{GeV}$\footnote{Larger values for the scalar mass splitting are highly constrained by oblique corrections ($S$,$T$,$U$ parameters).}, which corresponds to a very small $\lambda_5$. In this limit, the observed neutrino masses require relatively large Yukawa couplings $y_{\alpha k}$ (see Eq.~\eqref{eq:non_freePotentialParameters}).
\begin{figure}[!htbp]
    \def\sepf{0.496}
    \centering
    \includegraphics[width=\sepf\columnwidth]{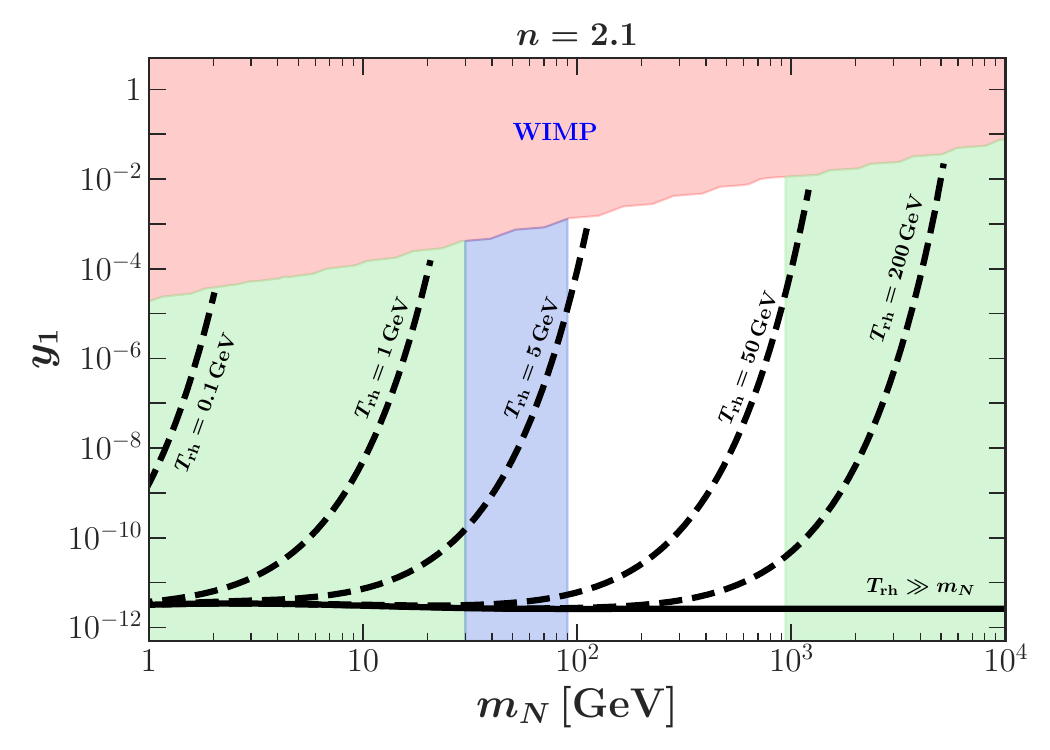}
    \includegraphics[width=\sepf\columnwidth]{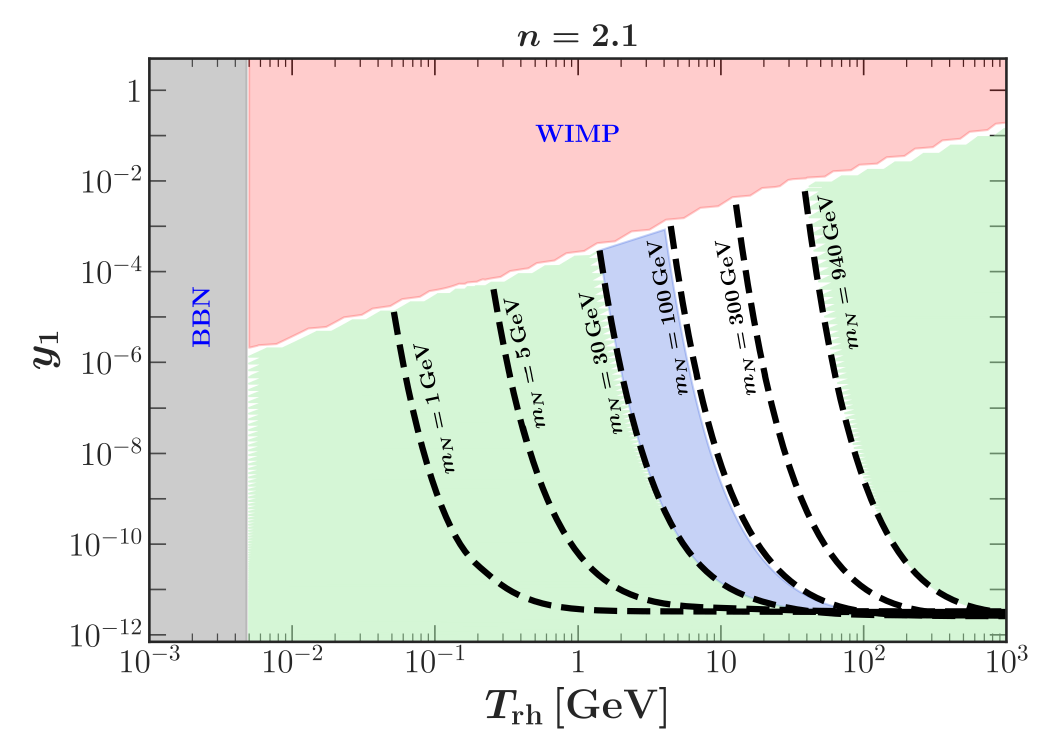}
   \caption{Parameter space required to reproduce the observed DM relic abundance for $n=2.1$, shown as a function of $m_N$ (left panel) and $T_{\rm rh}$ (right panel). BBN excludes the gray-shaded region in the right panel and corresponds to $T_{\rm rh} < T_{\rm BBN}$. The red-hatched region denotes the parameter space where $N$ reaches thermal equilibrium (WIMP regime). Oblique parameters exclude the blue region, and the white region represents the viable parameter space compatible with all model constraints.}
    \label{fig:n2}
\end{figure} 

\begin{figure}[!htbp]
    \def\sepf{0.496}
    \centering
    \includegraphics[width=\sepf\columnwidth]{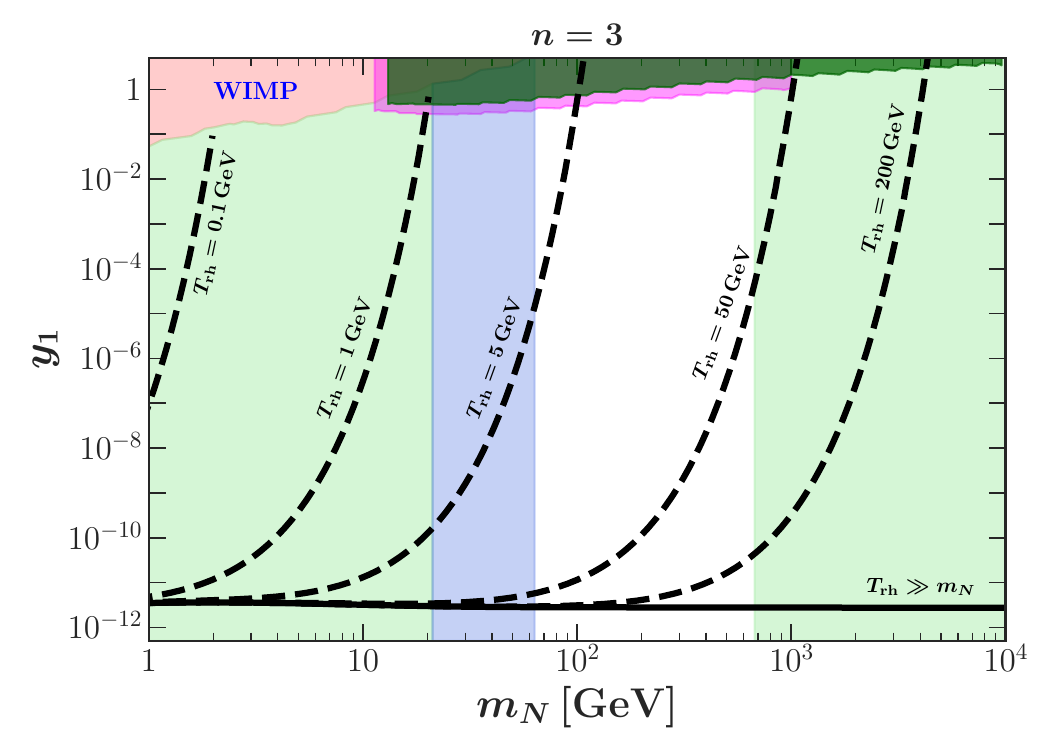}
    \includegraphics[width=\sepf\columnwidth]{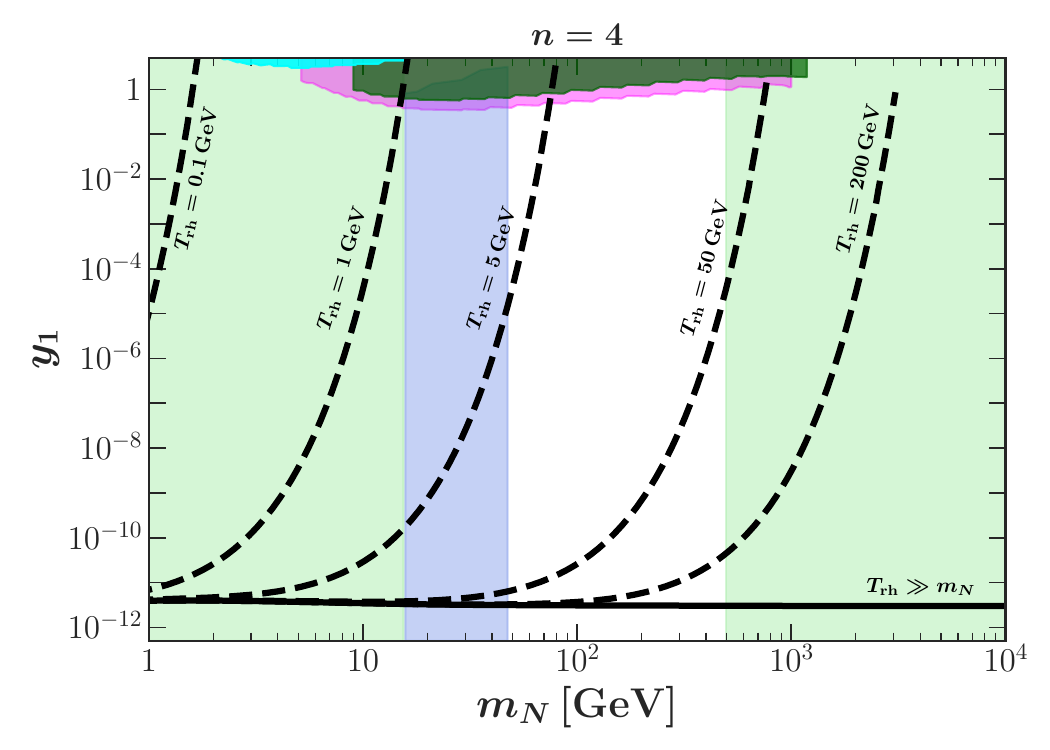}
    \caption{Parameter space required to reproduce the observed DM relic abundance for $n=3$ (left panel) and $n=4$ (right panel). The red-hatched region corresponds to the parameter space where $N$ reaches thermal equilibrium (WIMP regime). Oblique parameters exclude the blue region, and the white region represents the viable parameter space compatible with all model constraints. The green region indicates the parameter space excluded by LUX-ZEPLIN~\cite{LZ:2024zvo}, while the magenta region shows the projected sensitivity~\cite {DARWIN:2016hyl}. In the right panel, the orange region represents the projected sensitivity of Oscura~\cite{Oscura:2022vmi}.}
    \label{fig:n3}
\end{figure} 

Figs.~\ref{fig:n2} and~\ref{fig:n3}. This region is constrained from above by perturbativity and unitarity bounds. As the mass splitting between the charged and neutral scalars increases, the scalar couplings $\lambda_3$ and $\lambda_4$ exceed the perturbativity limit of $4\pi$ for dark matter masses $m_N \sim 1\text{ TeV}$. On the lower end, collider searches exclude the region where $m_N \lesssim 11\text{ GeV}$. These excluded regions are highlighted in light green in both figures. Additionally, the blue region is in tension with oblique parameters because the mass splitting between the charged and neutral scalars increases the contribution to the $T$ parameter (we checked this for large $\delta$ values).
The solid black curve shows the standard high-reheating regime ($T_{\rm rh} \gg m_N$), whereas the dashed black curves represent the low-reheating scenarios ($T_{\rm rh} \lesssim m_N$), which are among the central results of this work.

For the mild hierarchy $n=2.1$ shown in~\autoref{fig:n2}, the DM particle never reaches thermal equilibrium for phenomenologically relevant Yukawa couplings, so the freeze-in condition is automatically satisfied. Consequently, the Yukawa couplings remain too small to produce observable direct-detection signals. As the hierarchy increases, the thermal condition changes substantially. The heavier inert scalars become increasingly Boltzmann suppressed, reducing the production rate of $N$ and preventing thermalization over a much wider range of Yukawa couplings. As a consequence, values as large as $\mathcal{O}(0.1)$ remain compatible with the freeze-in mechanism, considerably enhancing the direct-detection prospects. This behavior is clearly visible in~\autoref{fig:n3} for $n=3$ and becomes even more pronounced for $n=4$.

Current LUX-ZEPLIN limits~\cite{LZ:2024zvo}, shown in green, already exclude part of the parameter space. The projected DARWIN sensitivity \cite{DARWIN:2016hyl}, shown in magenta, extends this reach over a large fraction of the region compatible with low reheating temperatures. In particular, for $n=3$,  DM masses in the range $60\,\mathrm{GeV}\lesssim m_N\lesssim700\,\mathrm{GeV}$ and reheating temperatures $5\,\mathrm{GeV}\lesssim \Trh\lesssim50\,\mathrm{GeV}$ are expected to be highly accessible to next-generation direct-detection experiments. For the largest hierarchy considered, $n=4$, DM-electron scattering also provides complementary sensitivity. 

Finally, as a complementary part of this work, we explore the thermal case in which $N$ acts as DM in the superWIMP scenario. As discussed in \autoref{sec:superWIMP}, $N_2$ can freeze out while remaining in thermal equilibrium with the SM plasma and subsequently decay completely into $N$ before BBN, thereby setting the $N$ relic abundance. To study this mechanism, we perform a parameter scan over the following ranges: $1\,\text{GeV} \le m_{N} \le 10^3\,\text{GeV}$, $1\,\text{GeV} \le m_{N_{2,3}} \le 10^4\,\text{GeV}$, $10^{-12} \le y_1 \le 1$, and $10^{-12} \le y_3 \le 10^{-8}$, with the scalar sector parameters set according to Eq.~\eqref{eq:scan-parameters}. We determine the coupling $y_2$ via Eq.~\eqref{eq:n2ton1ll} and Eq.~\eqref{eq:y1y2_BBN} to satisfy BBN limits, ensure $N_2$ freeze-out, and allow its subsequent decay into $N$.
To compute the thermal relic density of the NLOP, the $N_2$ field, we implement the model in SARAH~\cite{Staub:2008uz,Staub:2009bi,Staub:2010jh,Staub:2012pb,Staub:2013tta,Staub:2015kfa} and SPheno~\cite{Porod:2003um,Porod:2011nf} and export the model files to micrOMEGAs~\cite{Alguero:2023zol,Belanger:2026asz}. This code accounts for all relevant freeze-out processes, including resonances and coannihilations, in the Boltzmann-equation evolution in the early Universe.
\begin{figure}[h]
    \centering
    \includegraphics[width=0.55\linewidth]{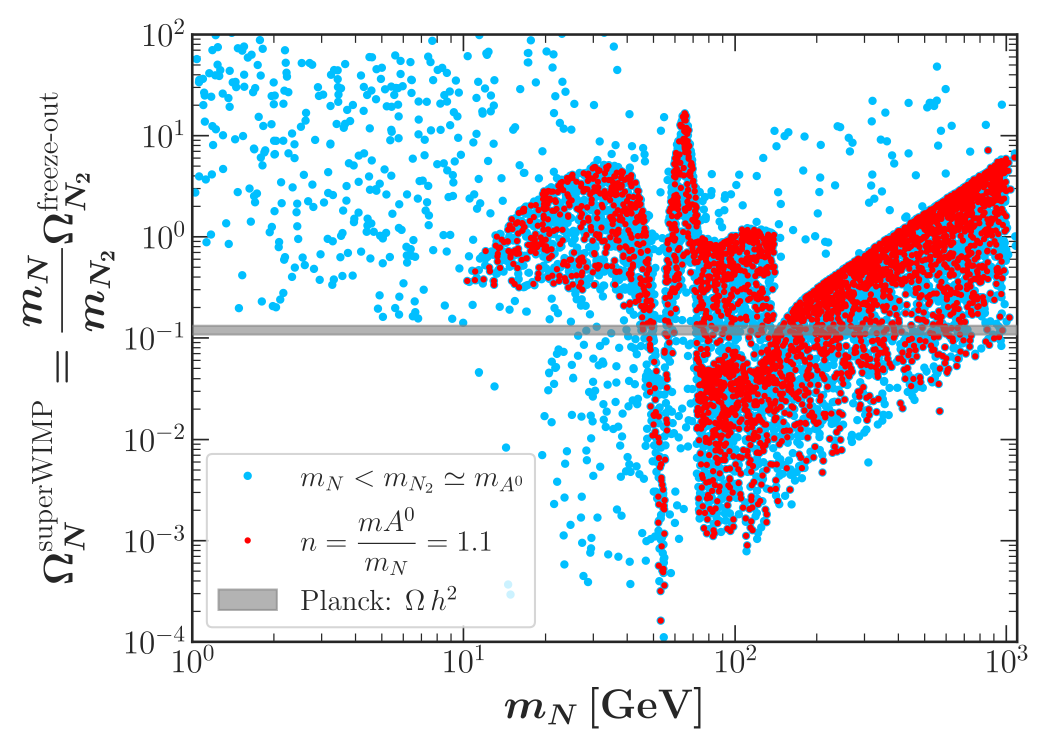}
    \caption{SuperWIMP relic density of $N$ as a function of its mass. The horizontal gray line marks the Planck-observed DM relic density~\cite{Planck:2018vyg}. Blue points highlight the region dominated by coannihilation processes, while red points denote a compressed spectrum with $n=1.1$ and cold DM.}
    \label{fig:superWIMP2}
\end{figure}

\autoref{fig:superWIMP2} presents the resulting $N$ relic abundance in the superWIMP regime. In the sub-GeV mass region, we always obtain an overabundance of DM because $N_2$ is an $\mathrm{SU(2)_L}$ singlet, leading to inefficient annihilation channels. Blue points represent the parameter space region corresponding to lower relic densities achieved through coannihilations between $N_2$ and the $A^0$ ($H^0$) scalar fields~\cite{Molinaro:2014lfa} $\left(m_{N} < m_{N_2} \simeq m_{A^0}, \ m_{H^0}\right)$. 
Red points depict the scenario with $n = 1.1$ in Eq.~\eqref{eq:scan-parameters}, which naturally fall within the coannihilation region. Since a superWIMP realization is obtained through a three-body decay $N_2\to N\,l\bar{l}$, we need low $n$ values to ensure a compressed spectrum and to have nonrelativistic DM at the time of structure formation (cold DM); otherwise, large $n$ values would trigger a hot DM and spoil the model~\cite{Gelmini_2010}.
\begin{figure}
    \centering
    \includegraphics[width=0.55\linewidth]{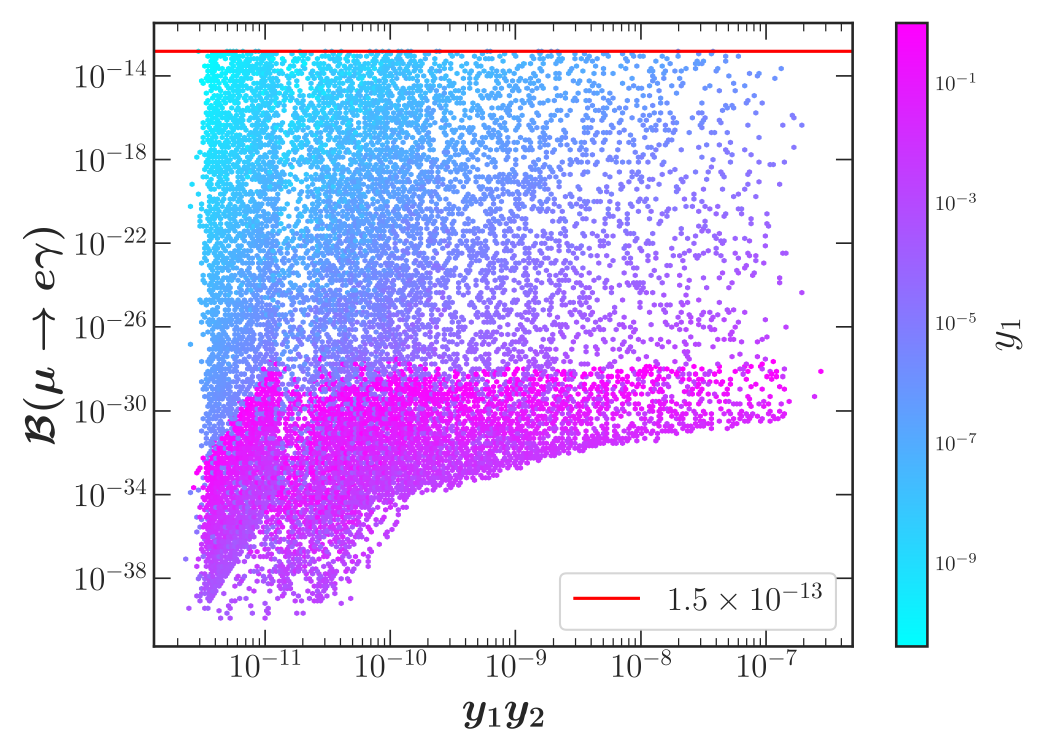}
    \caption{Contribution to the branching ratio $\mathcal{B}(\mu \to e \gamma)$ as a function of the product $y_1 y_2$. The color code represents the values of the Yukawa coupling $y_1$.}
    \label{fig:superWIMP4}
\end{figure}

As in previous sections, we also incorporate constraints from CLFV processes, particularly the decay $\mu \to e \gamma$, which is subject to the stringent experimental upper bound $\mathcal{B}(\mu \to e \gamma) < 1.5 \times 10^{-13}$~\cite{MEGII:2025gzr,SINDRUM:1987nra}.
When assuming a large Yukawa coupling $y_1$ to compensate for the Boltzmann suppression induced by a low reheating temperature, sizable contributions to the branching ratio $\mathcal{B}(\mu \to e \gamma)$ can be generated, thereby spoiling the viable parameter space of the model; however, we can control the Yukawa coupling $y_2$ and satisfy this current upper bound.
In~\autoref{fig:superWIMP4}, we show this branching ratio as a function of the product $y_1 y_2$, which is required to realize the superWIMP scenario discussed in \autoref{sec:superWIMP}.
Notice that $y_1 y_2$ has a lower bound of $\sim 10^{-12}$, dictated by the BBN constraint in \autoref{eq:y1y2_BBN}.
Furthermore, the plot demonstrates an upper bound of $y_1 y_2 \sim 10^{-6}$, required to guarantee the freeze-out of $N_2$ and its subsequent decay into $N_1$ (see \autoref{eq:n2ton1ll}), thereby satisfying all necessary conditions for the superWIMP mechanism.
In addition, Yukawa coupling $y_1 \sim \mathcal{O}(0.1)$ remains allowed by CLFV constraints, provided that $y_2$ is sufficiently small.

\section{Conclusions}
\label{sec:Conclusions}

In this work, we analyze the phenomenological implications of low reheating temperatures $T_{\rm rh}$ in the parameter space of the scotogenic model with the singlet fermion $N_1 = N$ as a FIMP DM candidate. The analysis shows that lowering the reheating temperature substantially enlarges the viable freeze-in parameter space and improves direct-detection prospects, with the details primarily governed by the mass hierarchy between the dark matter candidate and the lightest scalar, parameterized by the ratio $n = m_{A^0}/m_N$. We show that this mass hierarchy plays a critical role in thermalization and direct-detection searches for $N$. A tiny mass splitting between CP-odd and CP-even scalars, $\delta = 10^{-5}\,\text{GeV}$, allows relatively large Yukawa couplings because the yield is strongly Boltzmann-suppressed. This suppression prevents the Majorana fermion $N$ from reaching thermal equilibrium over a large portion of the parameter space while preserving the observed relic abundance. Consequently, broad regions of the parameter space corresponding to low reheating temperatures, $5\,\text{GeV} \lesssim T_{\rm rh} \lesssim 50\,\text{GeV}$ for masses in the range $60\,\text{GeV} \lesssim m_N \lesssim 700\,\text{GeV}$, become accessible to direct-detection experiments. In contrast, the transition to a large CP-odd and CP-even scalar mass splitting is strongly constrained by oblique corrections. Nevertheless, our results indicate that the increased DM number density at lower mass scales remains a strong prospect for direct detection. Both current LUX-ZEPLIN limits and future DARWIN projections can probe a significant fraction of this low-reheating parameter space, particularly for Yukawa couplings of $\mathcal{O}(0.1)$, while maintaining the model consistent with cLFV bounds. These results demonstrate that low reheating temperatures can reconcile the freeze-in paradigm with experimentally testable signatures, opening a viable and predictive region of the scotogenic parameter space that the next generation of direct-detection experiments will extensively explore.

\section{Acknowledgements}

We thank Oscar Zapata for insightful discussions. The work of R. L. is supported by ITM through the ``Convocatoria permanente para proyectos de investigación, I+D, I+D+i o I+C en modalidad recurso instalado de los grupos de investigación del ITM''. The work of D. S. is supported by Simons Foundation (Award Number:1023171-RC), FAPESP Grants 2021/01089-1, 2023/01197-4, ICTP-SAIFR FAPESP Grants 2021/14335-0, 
CNPq grants 403521/2024-6, 408295/2021-0, 403521/2024-6, 406919/2025-9, 351851/2025-9, ANID-Millennium Science Initiative Program ICN2019\_044, and IIF-FINEP grant 213/2024. 
The work of A. R. is supported by Sostenibilidad UdeA, CODI UdeA Grant No. 2024-76711.

\bibliographystyle{JHEP}
\bibliography{references}

\end{document}